\documentclass{article}
\usepackage[utf8]{inputenc}
\usepackage[T1]{fontenc}
\usepackage{amsmath}
\usepackage{amsfonts}
\usepackage{amssymb}
\usepackage{graphicx}
\usepackage{hyperref}
\usepackage{placeins}
\begin{document}
	\date{}
	\title{Spherically Symmetric Gravitational Collapse Of Inhomogeneous Dust Cloud In The Background Of Dark Energy}
	\maketitle
	\begin{center}
		\author{Anjali Pandey \footnote{anjalipandey903@gmail.com\, },{Rajesh Kumar \footnote{rkmath09@gmail.com\, },{Sudhir Kumar Srivastava \footnote{sudhirpr66@rediffmail.com\,}}\\
		\textit{Department of Mathematics $\&$ Statistics, \\
					Deen Dayal Upadhyaya Gorakhpur University, Gorakhpur, INDIA.}}\\}
	\end{center}		
	\begin{center}
		\textbf{Abstract}
	\end{center}
The	present paper deals with the gravitational collapse of an inhomogeneous spherical star consisting of  dust fluid ($\rho$) in the background of dark energy components ($p_{_{DE}},\rho_{_{DE}}$)  with equation of state $p_{_{DE}}=\omega \rho_{_{DE}}$ ($\omega$ being a non zero constant parameter describing dark energy component). We discussed the development of apparent horizon to investigate the black- hole formation in gravitational collapsing process. The collapsing process is examined first separately for dust cloud and dark energy and then under the combined effect of dust interactig with dark energy. It is obtained that when only dust cloud or dark energy is present the collapse leads to the formation of black-hole under certian conditions. When both of them are present, collapsing star does not form black-hole. However when dark energy is considered as cosmological constant $(\omega=-1)$, the collapse leads to black hole formation.\\

	\section{\textbf{INTRODUCTION}}\label{sec 1}
	Gravitational collapse is one of the most significant phenomenon in general theory of relativity. A massive star undergoes a continual gravitational collapse when the pressures inside it become insufficient to balance the pull of its own gravity. The problem of general relativistic gravitational collapse of massive stars has attracted  attention of researchers for many years, starting with the seminal paper by Oppenheimer and Snyder\cite{aa}. The motivation for such interest is easily understood-the gravitational collapse of massive stars represents one of the few observable phenomena, where general relativity is expected to play a relevant role and the physics of it being crucial and interesting amongst astrophysicists. The studies concentrated on the singularity formation (Black hole or Naked singularity) in gravitational collapse within the framework of general relativity and other gravitational theories are the interesting problems for theoretical astrophysicists\cite{bb}-\cite{hh}.  
	The astrophysical observation suggested that present universe consists of approximately 71 \% of dark energy and 15\% dark matter. The nature of dark energy as well as dark matter is unknown and many radically different models have been proposed, such as cosmological constant, quintessence\cite{ii,jj}, DGP branes \cite{kk, ll}, Gauss- Bonnet \cite{mm} and dark energy in brane worlds\cite{nn}-\cite{qq} for their studies. Such kinds of exotic fluids do not appear to interact with the usual standard model particles and therefore it is also very difficult to detect them by the modern detectors \cite{rr,ss}.\\
	 For the stars made of baryonic matter, besides gravitational interaction, they can also interact with each other by means of strong, weak and electromagnetic forces. Dark energy is supposed to interact with ordinary matter only through gravity. Therefore, only by its gravitational effects, one can gather information, such as its positions, mass, density profiles, etc. 
	A natural question is how dark energy affects the process of the gravitational collapse of a star. It is known that dark energy exerts a repulsive force on its surrounding and this repulsive force may prevent the star from collapse. Indeed, there are speculations that a massive stars do not simply collapse to form a black hole, instead to the formation of stars that contain dark energy. Some recent works have considered the spherically symmetric star consists of non-baryonic (DE and DM) matter and discussed its nature of singularity formation\cite{tt}-\cite{xx}. The dark energy, by virtue of its repulsive gravitational nature, is interesting to study gravitational collapse and formation of Black-hole. As all massive stars do not form black holes (may be neutron stars or white dwarfs), so it is generally seen \cite{ww,xx,yy} that the dark energy may play an important role in the collapsing stars. Gravitational collapse and formation of black holes in the presence of dark energy have been considered in several works \cite{xx,zz,ab}. Cai and Wang \cite{xx} investigated the black hole formation from homogeneous collapsing dust in the background of dark energy, and showed that the dark energy itself never collapses to form black holes but when both the dark energy and the dust are present, Black-Holes, can be formed due to the condensation of the dust. It has been shown that the mass of a Black-Hole decreases due to phantom energy accretion and tends to zero when the Big Rip approaches \cite{ac}. 
	Thus with the notion of such exotic content of universe, it is necessary to introduce this new kinds of matter distribution in our studies. Motivated from above works, we considered the inhomogeneous dust collapse in background of dark energy components and discuss the singularity formations. The paper is organised as- in section-2 we considered a class of shear-free spherically symmetric metric consisting of inhomogeneous dust fluid with dark energy. In sec.3 and 4 we discussed the gravitational collapse of an inhomogeneous dust cloud and dark energy separately. In section-5 we considered the combined effect of dust cloud interacting with dark energy and find that such interaction does not form the apparent horizon, that is, no formation of black hole. The last section-6 contains the conclusion part of the paper.

	\section{\textbf{Fluid distribution  and The field equations}}\label{sec2}
	Consider the collapsing spherically symmetric inhomogeneous star consists of dust cloud in the background of dark energy. The spherical system is  divide  in three different regions $\Sigma$, $ V^{+}$ and $ V^{-}$, where $\Sigma$ denotes the surface of the star, $ V^{+}$ and  
	$ V^{-}$ are exterior and the interior of the star.\\
	 The spherically symmetric inhomogeneous space time inside the star ( $ V^{-}$) is described by the metric
	\begin{equation} \label{eqn:1}
		ds^{2}_{-} = A^{2}(t,r)dt^{2} - C^{2}(t,r)(dr^{2} +d\Omega^{2})
	\end{equation} 
	where, $d\Omega^{2}=d\theta ^{2}+sin^{2}\theta d\phi^{2}$.
	The coordinate is taken as $x^{-i} = (t,r ,\theta ,\phi), i= 0,1,2,3$

	The energy momentum tensor  $T^{-}_{ij}$ of matter field inside the star is given by
	
	\begin{equation}\label{eqn:2}
		T^{-}_{ij}=(\rho + \rho_{_{DE}} + p_{_{DE}})u_{i}u_{j}-p_{_{DE}}g_{ij}
	\end{equation}
	
	where $\rho$ is energy density of inhomogeneous dust cloud, $\rho_{_{DE}}$ and $p_{_{DE}}$ are the energy density and pressure of dark energy respectively, while $u^{-}_{i}$ is four velocity vector given by $u_{i}=A(t,r)\delta^{0}_{i}$\\	
	The explicit form of the energy conservation equation $T^{-i}_{j;i}=0$ are 
	\begin{equation}\label{eqn:3}
		3(\rho + \rho_{_{DE}} + p_{_{DE}})\frac{\dot{C}}{C} +  (\dot{\rho}+\dot{\rho}_{_{DE}})=0
	\end{equation}
	\begin{equation}\label{eqn:4}
		(\rho+\rho_{_{DE}} + p_{_{DE}})\frac{A'}{A} + p_{_{DE}}'=0 
	\end{equation}
	where a dot (.) represents differentiation with respect to time $t$ and prime $(')$ represents differentiation with respect to $r$.\\
	The non vanishing component of Einstein field equation $(G_{ij}^{-}= - kT^{-}_{ij})$ for the metric (\ref{eqn:1}) and the matter field  of the form of equ.(\ref{eqn:2}) are given as, 
	\begin{equation}\label{eqn:5}
			G^{-}_{00} \equiv \frac{1}{C^2}+\frac{C'^2}{C^4}-2\frac{C''}{C^3}+3\frac{\dot{C}^2}{A^2C^2}=-k(\rho+\rho_{_{DE}})
		\end{equation}
	\begin{equation}\label{eqn:6}
 G^{-}_{11}\equiv\frac{2}{AC^3}A'C'+\frac{C'^2}{C^4}+\frac{2}{A^3C}\dot{A}\dot{C}-\frac{\dot{C}^2}{A^2C^2}-\frac{2}{A^2C}\ddot{C}-\frac{1}{C^2}= -kp_{_{DE}}
	\end{equation}
	\begin{equation}\label{eqn:7}
		G^{-}_{22}\equiv\frac{A''}{AC^2}-\frac{C'^2}{C^4}+\frac{C''}{C^3}+
		+\frac{2}{A^3C}\dot{A}\dot{C}-\frac{\dot{C}^2}{A^2C^2}-2\frac{\ddot{C}}{A^2C}=-kp_{_{DE}}
	\end{equation}
	\begin{equation}\label{eqn:8}
		G^{-}_{33}=  \textrm {sin}\theta G_{22}
	\end{equation}
	\begin{equation}\label{eqn:9}
		G^{-}_{01}=G^{-}_{10}\equiv\frac{A'}{A}+\frac{C'}{C}-\frac{\dot{C'}}{\dot{C}}=0
	\end{equation}
	where $k=\frac{8\pi G}{c^4}$.
	An apparent horizon is described as the boundary of all trapped surfaces and the black hole formation is identified by the development of apparent horizon on which we have
	\begin{equation}\label{eqn:10}
		C,_{\alpha}C,_{\beta}g^{-\alpha\beta}= \frac{\dot{C^2}}{A^2}	-\frac{C'^2}{C^2} = 0	
	\end{equation}	
	
	where$ (),_{x} \equiv \partial()/\partial x$\\
	Another important quantity to describe the collapse is the mass function $m(t,r)$, given by 
	\begin{equation}\label{eqn:11}
		m(t,r)  =\frac{1}{2}C\left(1+\frac{\dot{C^2}}{A^2}-\frac{C'^2}{C^2}\right)
	\end{equation}
	which can be intepreted  mass inside the collapsing star any instant $(t,r)$.\\ 
	Assuming that $t_{AH}$ be the time when the whole star is collapses inside the apparent horizon surface $r= r_{_{AH}}$, then equ.(\ref{eqn:10}) gives 
	\begin{equation}\label{eqn:12}
		\frac{\dot{C}^2(t_{_{AH}}, r_{_{AH}})}{A^2(t_{_{AH}}, r_{_{AH}})}	= \frac{C'^2(t_{_{AH}}, r_{_{AH}})}{C^2(t_{_{AH}} r_{_{AH}})} 
	\end{equation}
	 The total mass contributed by the collapsing star to the mass of black hole is 
	\begin{equation}\label{eqn:13}
		M_{_{BH}}= M(t_{_{AH}})
	\end{equation}	
	If no matter continously falls into the black hole from outside of the star after the formation of apparent horizon (i.e after the moment $t_{_{AH}}$), we can calculate the total mass of the collapsing body (as the total mass of the black hole) by above equ.(\ref{eqn:13}).
	 We also assume that the collapsing star is not initially trapped then we should have
	\begin{equation}\label{eqn:14}
		\frac{\dot{C^2}}{A^2}	-\frac{C'^2}{C^2} < 0	
	\end{equation}

	\section{\textbf{Gravitational Collapse of Dust Cloud}}\label{sec3}
	In this case we assume that
	\begin{equation}\label{eqn:15}
		\rho \neq 0 , \rho_{_{DE}} = 0 = p_{_{DE}}
	\end{equation}
	that is, the collapsing star consist of a only dust cloud. Historically, gravitational collapse of dust cloud studied by Openheimer and Synder\cite{aa}, which leads to the formation of black holes. In this section we adapted the different approach to discuss the singularity formation of collapsing dust cloud.\\
	In view of equs.(\ref{eqn:3}) and (\ref{eqn:15}) we find that
	\begin{equation}\label{eqn:16}
	 \rho=\frac{\rho_0}{C^3}
	 	\end{equation}
	  where $\rho_0 \equiv \rho_0(r)$ is constant of integration.
	  It follows from  equs.(\ref{eqn:4}) and (\ref{eqn:15}) that $A(t,r) \equiv A(t)	$.
	For the suitable coordinate transformation $A(\tilde{t})d\tilde{t} \rightarrow dt$ one may take $A\sim 1$. Therefore equ.(\ref{eqn:9}) gives 
	\begin{equation}\label{eqn:17}
		C=\mu(t)\beta(r)
	\end{equation}	
	where $\mu(t)$  and $\beta(r)$  arbitarary functions choosen as integrating constants.\\
 Since for collapsing configuration $\frac{\dot{C}}{C}<0$  which implies $\dot{\mu}<0$ \\
Thus we have from equs.(\ref{eqn:16}) and (\ref{eqn:17})
\begin{equation}\label{eqn:18}
	\rho=\frac{\rho_0}{\beta^3\mu^3}
\end{equation}
\begin{equation}\label{eqn:19}
	\frac{\dot{C}}{C}=\frac{\dot{\mu}}{\mu}
\end{equation}	
	
Let us assume the star starts to collapse from $t=t_0$ , $r=r_{\Sigma}$ (since the star is supposed not initially trapped ie., satisfying equ(\ref{eqn:14})) and end at $t=t_e$ and $r=r_e$ (singularity formation). It can be seen from equs.(\ref{eqn:18}-\ref{eqn:19}) that collapse of star end when $\mu(t_e)$= constant= 0 where a spacetime singularity is finally formed $(\rho \rightarrow \infty, \dot{C}\rightarrow 0)$ \\
From equ.(\ref{eqn:10}), the development of apparent horizon in this case can be given by
\begin{equation}\label{eqn:20}
	\dot{\mu}= \frac{\beta'}{\beta^2}
\end{equation}
 Let apparent horizon is formed at the epoch $(t_{AH},r_{AH})$, then from equ.(\ref{eqn:20}) we have
\begin{equation}\label{eqn:21}
-K(t_e-t_{_{AH}}) = \mu(t_{_{AH}}) > 0
\end{equation}
and
\begin{equation}\label{eqn:22}
	K(r_e-r_{_{AH}})= \frac{\beta(r_e)-\beta(r_{_{AH}})}{\beta(r_e)\beta(r_{_{AH}})}
\end{equation}	
	where, 
	$K=\frac{\beta'}{\beta^2}<0$.
Thus, in this case the horizon is formed before the end-state $(t_{_{AH}}<t_e)$ and therefore the collapse will form black-hole.\\
It should be noted that if the collapse end as black-hole, then $r_e < r_{_{AH}}$. Therefore from equ.(\ref{eqn:22}) \\

$$\beta(r_{_{AH}})-\beta(r_e)<0$$
Thus in the present case the arbitrary function $\beta(r)$ may be choosen as $\frac{1}{r^n}$, n is a positive integer. Then the mass of the black hole from equs.(\ref{eqn:11}) and (\ref{eqn:13}) is 
\begin{equation}\label{eqn:23}
	M_{BH}= \frac{\mu(t_{_{AH}})}{2r_{_{AH}}^{3n}}\left[\dot{\mu}^2(t_{_{AH}})+r^{2n}_{_{AH}}(1-\frac{n^2}{r_{_{AH}}^2})\right]
\end{equation}

	\section{Gravitational Collapse of Dark energy}\label{sec4}
	In this section we discuss the gravitational collapse in the background of dark energy component only, 
	\begin{equation}\label{eqn:24}
		\rho = 0 ,\quad  p_{_{DE}} = \omega \rho_{_{DE}} \neq 0
	\end{equation}
	where the parameter $\omega (\neq -1)$ is a non zero constant.\\
Then equ.(\ref{eqn:9}) yields\\
	\begin{equation}\label{eqn:25}
		A= \nu(t)\frac{\dot{C}}{C}
	\end{equation} 
where $\nu(t)$ is arbitrary. 	
	In this case equ.(\ref{eqn:3}) yields
	\begin{equation}\label{eqn:26}
		\rho_{DE} = \frac{1}{C^{3(1+\omega)}}
	\end{equation}

	In view of equs.(\ref{eqn:25}) and (\ref{eqn:26}) we obtain from equ.(\ref{eqn:4}) 
	\begin{equation}\label{eqn:27}
		\frac{\dot{C}^{\frac{1+\omega}{\omega}}}{C^{\frac{1+4\omega+3\omega^2}{\omega}}}=\lambda(t)
	\end{equation}
where $\lambda $ is arbitrary \\
Let us assume that the collapse of star end at $(t_e,r_e)$ where a spacetime singularity is formed. It can also be seen from equ.(\ref{eqn:27}) that collapse end when $\lambda(t_e)\rightarrow 0$.\\
Consider the star is not initially trapped and let apparent horizon is formed at $(t,r)= (t_{_{AH}},r_{_{AH}})$.
Then, from equ(\ref{eqn:15}) development of apparent horizon gives 
\begin{equation}\label{eqn:28}
	\frac{C(r_{_{AH}},r_{_{AH}})-C(t_e,r_e)}{C(t_{_{AH}},r_{_{AH}}) C(t_e,r_e)}=\frac{\nu(t_e)r_{_{AH}}-\nu(t_{_{AH}})r_e}{\nu(t_{_{AH}})\nu(t_e)}
\end{equation}
 Further the effect of dark energy on collapsing star will be studied by parameterizing $\omega$.\\
\textbf{Case (i) $ \omega=-\frac{2}{3}$}

Soma et al.\cite{zz} obtained the singularity formation for $-1\leq \omega \leq -\frac{2}{3}$ and showed that when $\omega<-\frac{2}{3}$ collapse do not form Black-hole. However for $\omega= -\frac{2}{3}$ the end state of collapse may be either Black-hole or Nacked singularity depending on initial density.\\
In this case, it follows from equ.(\ref{eqn:27}) that 
\begin{equation}\label{eqn:29}
	C\dot{C}=\lambda^2(t)
\end{equation}
which gives by integrating
\begin{equation}\label{eqn:30}
	C=\sqrt{\lambda_1(t)+\beta(r)}
\end{equation}
where $\beta(r)$ and $\lambda_1(t)=2\int \lambda^2(t)dt$ are arbitrary. Also from equ.(\ref{eqn:26}) that
\begin{equation}\label{eqn:31}
	\rho_{DE}=\frac{1}{\sqrt{\lambda_1(t)+\beta(r))}}
\end{equation}

Now, in new of equs.(\ref{eqn:28}) and (\ref{eqn:30}) we observe an apparent horizon form (black hole formation) when
$$\lambda_1(t_{_{AH}})>\beta(r_e)-\beta(r_{_{AH}})$$

The contribution of collapsing star to the mass of black-hole,

	\begin{multline}\label{eqn:32}
 M_{_{BH}} = \frac{1}{2}[\lambda_1(t_{_{AH}})+\beta(r_{_{AH}})]^{3/2}
 \biggl[\frac{1}{\mu^2(t_{_{AH}})}+\\
 \frac{1}{4(\lambda_1(t_{_{AH}})+\beta(r_{_{AH}}))^3}
 \{4(\lambda_1(t_{_{AH}})+\beta(r_{_{AH}}))^2-\beta'^2(r_{_{AH}})\}\biggr]
 \end{multline}

\textbf{Case (ii) Phantom fluid}\\
The equation of state with $\omega<-1$ is named as " Phantom fluid" has recieved attention amongest researchers\cite{ad}\cite{ae}. For example $\omega=-\frac{2}{3}$, the phantom density increases with time which voilates the dominant energy condition.\\
Consider $\omega=-\frac{2}{3}$, we obtain from equ.(\ref{eqn:27}) that
\begin{equation}\label{eqn:33}
	C=[\lambda_2(t)+\alpha(r)]^{\frac{2}{9}}	
\end{equation} 
where $\alpha(r)$ and $\lambda_2(t)= \int \frac{9}{2}\lambda^3(t)dt$ are arbitrary.
From equ.(\ref{eqn:26})
\begin{equation}\label{eqn:34}
	\rho_{_{DE}}= [\lambda_2(t)+\alpha(r)]^{\frac{1}{3}}
\end{equation}

It follows from equs.(\ref{eqn:28}) and (\ref{eqn:33}) that the collapse end as black hole (formation of an apparent horizon) when
$$\lambda_2(t_{_{AH}})> \alpha(r_e)-\alpha(r_{_{AH}})$$ 
and the mass of the formed black hole is
\begin{equation}\label{eqn:35}
	\begin{split}
		M_{_{BH}}= \frac{1}{2}[\lambda_2(t_{_{AH}})+\alpha(r_{_{AH}})]^{\frac{-16}{9}}[\{1+(\lambda_2(t_{_{AH}})+\alpha(r_{_{AH}}))^{\frac{4}{9}}\}
		(\lambda_2(t_{_{AH}})+\alpha(r_{_{AH}}))^2-\\
		\frac{4}{81}\alpha'^2]
	\end{split}
\end{equation}

\section{Gravitational Collapse with combined effect of dust cloud and dark energy}

Recently, number of authors have been considered the interaction of dust cloud and dark energy by assuming(\cite{uu,xx,ab,am} and references therein)
\begin{equation}\label{eqn:36}
\frac{\rho_{_{DE}}}{\rho}=\gamma  C^{3m}
\end{equation}
  
where $\lambda$(>0) and $m$ are arbitrary constants and $C(t,r)$ be the geometric radius of star. The relation (\ref{eqn:36}) shows that the energy flow between dust fluid and dark energy. Assuming that the dark energy satisfy the equation of state $p_{_{DE}}=\omega \rho_{_{DE}}$, $\omega$ being a constant parameter.\\
From equ.(\ref{eqn:3}),(\ref{eqn:9}) and (\ref{eqn:36}) we obtain

\begin{equation}\label{eqn:37}
	\rho_{_{DE}}=C^{-3(1-m)}(1+\gamma C^{3m})^{-\frac{m+\omega}{m}}
\end{equation}

\begin{equation}\label{eqn:38}
	\rho=\frac{C^{-3}}{\gamma}(1+\gamma C^{3m})^{-\frac{m+\omega}{m}}
\end{equation}

It follows from above equations that the spacetime has central singularity  at $C(t,r)=0$. Taking use of eqns. $(\ref{eqn:37})-(\ref{eqn:38})$ into equ.(\ref{eqn:4}), we obtain
\begin{equation}\label{eqn:39}
\begin{split}
\dot{C'}(1+\l\gamma C^{3m})[1+\gamma(1+\omega)C^{3m}]-\dot{C}C'[1+\gamma^2(1+3\omega^2+4\omega)C^{3m}\\
-\gamma\{-2+(3m-4)\omega\}]C^{3m-1}=0
\end{split}
\end{equation}

The equ.(\ref{eqn:39}) is highly non-linear in $C(t,r)$ therefore for further discussion in the following sections we consider different value of $m$ and $\omega$ to investigate the collapsing phenomenon and apparent horizon formation of dust cloud interacting with dark energy.

\begin{center}
	\textbf{(A).\quad m = $\frac{1}{2}$ }
\end{center}
when $m=\frac{1}{2}$, it follows from equ.(\ref{eqn:39}) that
\begin{equation}\label{eqn:40}
	\frac{\dot{C}}{C}= \begin{cases}
		\frac{\beta_1(t)}{(1+\gamma C^{\frac{3}{2}})}, & \text{for $\omega = -1$} \\
		\frac{\beta_2(t)}{(1+\gamma C^{\frac{3}{2}})(3+\gamma C^{\frac{3}{2}})}, & \text{for $\omega = -\frac{2}{3}$ }\\
		\frac{\beta_3(t)}{(1+\gamma C^{\frac{3}{2}})^2(2-\gamma C^{\frac{3}{2}})}, & \text{for $\omega = -\frac{3}{2}$}
	\end{cases}
\end{equation}

where $\beta_1, \beta_2$ and $\beta_3$ are arbitrary functions of $t$ and 
\begin{equation}\label{eqn:41}
	C e^{\frac{2}{3}\gamma C^{\frac{3}{2}}}= \alpha_1(t)K_1(r) \quad   \textrm{for $\omega$= -1}
\end{equation}
 
\begin{equation}\label{eqn:42}
	\begin{split}
	13+14 \gamma C^{\frac{3}{2}}+ \gamma^2 C^3 -\frac{6}{\gamma c^{\frac{3}{2}}} (1+\gamma C^{\frac{3}{2}})^{\frac{2}{3}}  \textrm{Hypergeometric} 2F1\left[\frac{2}{3},\frac{2}{3},\frac{5}{3},-\frac{1}{\gamma C^{\frac{3}{2}}}\right]\\
=	2[\alpha_2(t)+ K_2(r)](1+\gamma C^{\frac{3}{2}})^{\frac{2}{3}} \quad \textrm{for $\omega=-\frac{2}{3}$}
\end{split}
\end{equation}
 and
\begin{equation}\label{eqn:43}
	C e^{\gamma C^{\frac{3}{2}}}-\frac{1}{9}\gamma^3C^{\frac{9}{2}}= K_3(r)\alpha_3(t) \quad \textrm{for $\omega= -\frac{3}{2}$} 
\end{equation}

where $\alpha_1(t) , \alpha_2(t) , \alpha_3(t), K_1(r), K_2(r)$, and $K_3(r)$
are arbitrary constants of integration. We can observe from equ.(\ref{eqn:40}) the collapsing process will not end untill the $\beta_1, \beta_2, \beta_3$ are zero.\\
Also from equs.(\ref{eqn:37}) and (\ref{eqn:38}) we have
\begin{equation}\label{eqn:44}
	\rho_{_{DE}} = \begin{cases}
		\gamma + \frac{1}{C^{\frac{3}{2}}}, & \text{for $\omega= -1$}.\\
		\frac{(1+\gamma C^{\frac{3}{2}})^{\frac{1}{3}}}{C^{\frac{3}{2}}}, &\text{for $\omega=- \frac{2}{3}$}\\
	\frac{(1+\gamma C^{\frac{3}{2}})^2}{C^{\frac{3}{2}}}, & \text{for $\omega =-\frac{3}{2}$}	
\end{cases}
\end{equation}

\begin{equation}\label{eqn:45}
	\rho = \begin{cases}
		
	\frac{(1+\gamma C^{\frac{3}{2}})}{C^3}	, & \text{for $\omega- -1$}\\
	\frac{(1+\gamma C^{\frac{3}{2}})^{\frac{1}{3}}}{C^3}, & \text{for $\omega= -\frac{2}{3}$}.\\
	\frac{(1+\gamma C^{\frac{3}{2}})^2}{C^3}, & \text{for $\omega= -\frac{3}{2}$}
					
	\end{cases}
\end{equation}

The development of apparent horizon in this case are,
\begin{equation}\label{eqn:46}
	C_{_{AH}}(1+\gamma C_{_{AH}}^{\frac{3}{2}})= K_{1}(r_{_{AH}})\mu(t_{_{AH}}) \quad  \textrm {for $\omega$= -1}
\end{equation}

\begin{equation}\label{eqn:47}
C_{_{AH}}(1+\gamma C_{_{AH}}^{\frac{3}{2}})^{\frac{1}{3}}(3+\gamma C_{_{AH}}^{\frac{3}{2}})= K_{2}(r_{_{AH}})\mu(t_{_{AH}}) \quad  \textrm{for $\omega= -\frac{2}{3}$}
\end{equation}
 and

\begin{equation}\label{eqn:48}
	C_{_{AH}}(1+\gamma C_{_{AH}}^{\frac{3}{2}})^2 (2-\gamma C_{_{AH}}^{\frac{3}{2}})= K_3(r_{_{AH}})\mu(t_{_{AH}}) \quad \textrm{for $\omega= -\frac{3}{2}$}
\end{equation}

where $C_{_{AH}}$ being surface radius of apparent horizon at instant $(t_{_{AH}}, r_{_{AH}})$. The equations (\ref{eqn:46})-(\ref{eqn:48}) have no real solution for $C_{_{AH}}$, therefore the collapse denied to the formation of apparent horizon surface.
It also follows from equs.(\ref{eqn:44})-(\ref{eqn:45}) that the collapsing star has only central singularity when $ C(t,0)=0$.

$$\textbf{(B).\qquad m=1}$$
In this case For $\omega=-1$, eqn.(\ref{eqn:39}) yields\\
when $\omega=-1$\\
\begin{equation}\label{eqn:49}
	C= \eta_1(t) \xi_1(r)
\end{equation}
\begin{equation}\label{eqn:50}
	\frac{\dot{C}}{C}= \frac{\dot{\eta_1}}{\eta_1}
\end{equation}
\begin{equation}\label{eqn:51}
	\rho = \frac{1}{\eta_1^3\xi_1^3}, \quad \rho_{_{DE}}=1
\end{equation}
where $\eta_1(t)$ and $\xi_1(r)$ are arbitrary constants of integration.
The development of apparent horizon in this case is
\begin{equation}\label{eqn:52}
	C_{_{AH}}= \xi_1(r_{_{AH}}) \mu(t_{_{AH}})
\end{equation}

From equs.(\ref{eqn:50})-(\ref{eqn:51}) we see that the singularity occured when either $\eta_1(t)=0$ or $\xi_1(r)=0$ or both zero and equ.(\ref{eqn:52}) reveals the formation of apparent horizon.\\ 
When $\omega= -\frac{2}{3}$ and $\omega=-\frac{3}{2}$, the equ.(\ref{eqn:39}) yields 
\begin{equation}\label{eqn:53}
	\frac{\dot{C}}{C}= \begin{cases}
		\frac{\eta_2(t) (1+\gamma C^3)^{\frac{1}{3}}}{(3+\gamma C^3)}, & \text{for $\omega= -\frac{2}{3}$}.\\
		\frac{\eta_3(t) (1+\gamma C^3)^{\frac{1}{2}}}{(2+\gamma C^3)}, & \text{for $\omega = -\frac{3}{2}$}
	\end{cases}
\end{equation}
where $\eta_2(t), \eta_3(t)$ are arbitrary and 

\begin{equation}\label{eqn:54}
	\begin{split}
	1+ \gamma C^3- \frac{6(1+\gamma C^3)^{\frac{1}{3}}}{(\gamma c^3)^{\frac{1}{3}}}   \textrm{Hypergeometric} 2F1\left[\frac{1}{3}, \frac{1}{3}, \frac{4}{3}, -\frac{1}{\gamma c^3}\right]-\\
	 2 (1+\gamma C^3)^{\frac{1}{3}} \eta_4(t)=0  \quad  \textrm{for}\quad \omega=-\frac{2}{3}
\end{split}
\end{equation}
 and
\begin{equation}\label{eqn:55}
	2\textrm{tanh}^{-1}(1+\gamma C^3)^{\frac{1}{2}}- (1+\gamma C^3)^{\frac{1}{2}} + \frac{3}{2} \eta_{5}(t)=0 \quad  \textrm{for $\omega= -\frac{3}{2}$}
\end{equation}

The development of of apparent horizon
\begin{equation}\label{eqn:56}
	C_{_{AH}}(3+\gamma C_{_{AH}}^3)- \xi_2(r_{_{AH}}) \mu(t_{_{AH}}) (1+\gamma C_{_{AH}}^3)^{\frac{1}{3}}=0 \quad  \textrm{for $\omega=-\frac{2}{3}$}
\end{equation}

\begin{equation}\label{eqn:57}
	C_{_{AH}}(2+\gamma C_{_{AH}}^3)- \xi_3(r_{_{AH}}) \mu(t_{_{AH}}) (1+\gamma C_{_{AH}}^3)^{\frac{1}{2}}=0 \quad  \textrm{for $\omega=-\frac{3}{2}$}
\end{equation}

here $C_{_{AH}}$ is the surface radius of apparent horizon at moment $(t_{_{AH}},r_{_{AH}})$ and $ \xi_2 , \xi_3$ are arbitrary. Since equs.(\ref{eqn:56})  and (\ref{eqn:57}) have no real solution, therefore  collapsing star do not form apparent horizon for $\omega= -\frac{2}{3}$ and$ -\frac{3}{2}$. It can  also be seen from equ.(\ref{eqn:53}) that collapsing process will not end untill $\eta_2$ and $\eta_3$ are zero.

\section{Conclusing Remarks}\label{5} 
Owing to the perplexing nature of dark energy, its  study got much interest in the last two decades(\cite{af,ag,ah} and references therein). Due to negative pressure, there are speculation that collapse with dark energy component should not form black hole. However, some recent works has been consider the gravitational collapse of  spherically symmetric homogeneous star in background of dark energy and dark matter and discuss their singularity formation (\cite{uu},\cite{xx},\cite{ai}-\cite{al}). In this paper,  we  have considered the gravitational collapse of an inhomogeneous spherical star consisted of dust cloud and dark enegry  and discuss the problem of black hole formation.We considered three distinct cases of gravitational collapse seperately- (i) collapse with dust cloud only (ii) collapse with dark energy only (iii) gravitational collapse with combined effect of dust and dark energy. We have discussed the  formation of black hole by the development of apparent horizon in collapsing process. The obtained results in gravitational collapse of dust cloud case manifest the formation of black hole which consistent with the oppenheimer and Synder result\cite{aa}. For dark energy in the form of perfect fluid has considered with the two candidate of dark energy components $\omega<-\frac{1}{3}$ and $\omega< -1$ (phantom). When $\omega=-\frac{2}{3}$, we observed the formation of apparent horizon (black hole) under the condition $\beta(r_e)-\beta(r_{_{AH}})<\lambda_1(t_{_{AH}})$. For $\omega=-\frac{3}{2}$, it is observed that due to presence of phantom fluid, the collapse end as black hole when $\lambda_2(t_{_{AH}})> \alpha(r_{e})-\alpha(r_{_{AH}})$. Thus we see that the collapsing stars which consist only dark energy fluid, may also leads to the Black-hole formation.\\
In section(5) we consider the combined effect of dust cloud and dark energy on the gravitational collapse. To study the effect of interaction between dust cloud and dark energy we consider the gravitational collapse by assuming that $\rho_{_{DE}}= \gamma \rho C^{3m}$, where $\gamma$ and $m$ are arbitrary constants\cite{xx,am} and equation of state $p_{_{DE}}=\omega \rho_{_{DE}}$, $\omega$ being non-zero constant parameters. By considering several specific models for different value of $m=\frac{1}{2},1$ and $\omega =-1, -\frac{2}{3}, -\frac{3}{2}$ we obtained the analogous conclusions that gravitational collapse denied the formation of apparent horizon, that is, black- holes do not form due to the collapse of dust cloud interacting with dark energy. Although when  $m=1$, $\omega=-1$ (cosmological constant) the collapse may interpret the development of apparent horizon. In ref.\cite{xx}\cite{an} several collapsing models on homogeneous and isotropic fluids have been discussed and found that black hole may be formed by the gravitational collapse even when considering the interaction between dust cloud and dark energy. Since the collapsing stars that consist of homogeneous and isotropic fluid are very ideal case, and in more realistic cases are the internal region of the collapsing star should be inhomogeneous. Therefore we believe that the results of the present paper will be in more realistic cases. However our results obtained in this paper do not seemingly support the speulations that black holes do not exists due to presence of dark energy.


\begin{thebibliography}{99}
\bibitem{aa} J Robert Oppenheimer and Hartland Snyder. On continued gravitational contraction. Physical Review, 56(5):455, 1939.
\bibitem{bb} SR Maiti. Fluid with heat flux in a conformally flat space-time. Physical
Review D, 25(10):2518, 1982.
\bibitem{cc} Bijan Modak. Cosmological solution with an energy flux. Journal of Astrophysics and Astronomy, 5(3):317–322, 1984.
\bibitem{dd} A Banerjee, SB Dutta Choudhury, and Bidyut K Bhui. Conformally flat
solution with heat flux. Physical Review D, 40(2):670, 1989.
\bibitem{ee} Dirk Schafer and Hubert F Goenner. The gravitational field of a radiating and contracting spherically-symmetric body with heat flow. General Relativity and Gravitation, 32(11):2119–2130, 2000.
\bibitem{ff} BVIvanov. Collapsing shear-free perfect fluid spheres with heat flow. General Relativity and Gravitation, 44(7):1835–1855, 2012.
\bibitem{gg} L Herrera, G Le Denmat, NO Santos, and A Wang. Shear-free radiating
collapse and conformal flatness. International Journal of Modern Physics
D, 13(04):583–592, 2004.
\bibitem{hh} Soumya Chakrabarti and Narayan Banerjee. Scalar field collapse in a conformally flat spacetime. The European Physical Journal C, 77(3):1–9, 2017.
\bibitem{ii} Andrew R Liddle and Robert J Scherrer. Classification of scalar field potentials with cosmological scaling solutions. Physical Review D, 59(2):023509, 1998.
\bibitem{jj} Paul J Steinhardt, Limin Wang, and Ivaylo Zlatev. Cosmological tracking solutions. Physical Review D, 59(12):123504, 1999.
\bibitem{kk} Gia Dvali, Gregory Gabadadze, and Massimo Porrati. 4d gravity on a brane in 5d minkowski space. Physics Letters B, 485(1-3):208–214, 2000.
\bibitem{ll} Cedric Deffayet. Cosmology on a brane in minkowski bulk. Physics Letters B, 502(1-4):199–208, 2001.
\bibitem{mm} Daniele Malafarina, Bobir Toshmatov, and Naresh Dadhich. Dust collapse in 4d Einstein–Gauss–Bonnet gravity. Physics of the Dark Universe, 30:100598, 2020.
\bibitem{nn} James M Cline and Jeremie Vinet. Problems with time-varying extra dimensions or “cardassian expansion” as alternatives to dark energy. Physical Review D, 68(2):025015, 2003.
\bibitem{oo} Yungui Gong, Anzhong Wang, and Qiang Wu. Cosmological constant and
late transient acceleration of the universe in the horava–witten heterotic m theory on s1/z2. Physics Letters B, 663(3):147–151, 2008.
\bibitem{pp} PR Pereira, MFA Da Silva, and R Chan. Anisotropic self-similar cosmological
model with dark energy. International Journal of Modern Physics D, 15(07):991–999, 2006.
\bibitem{qq} CFC Brandt, R Chan, MAF Da Silva, and Jaime F Villas Da Rocha. Inhomogeneous dark energy and cosmological acceleration. General Relativity and Gravitation, 39(10):1675–1687, 2007.
\bibitem{rr} Vera C Rubin, W Kent Ford Jr, and Norbert Thonnard, The Astrophysical Journal, 225:L107–L111, 1978.
\bibitem{ss} Vera C Rubin,WKent Ford Jr, and Norbert Thonnard. Rotational properties
of 21 sc galaxies with a large range of luminosities and radii, from ngc 4605/r= 4kpc/to ugc 2885/r= 122 kpc. The Astrophysical Journal, 238:471– 487, 1980.
\bibitem{tt} Prabir Rudra, Ritabrata Biswas, and Ujjal Debnath. Presence of dark energy
and dark matter: does cosmic acceleration signifies a weak gravitational
collapse? Astrophysics and Space Science, 342(2):557–574, 2012.
\bibitem{uu} Syed Zaheer Abbas, Hasrat Hussain Shah, Huafei Sun, Farook Rahaman, and Faizuddin Ahmed. Gravitational collapse of dust fluid and dark energy in
the presence of curvature: Black hole formation. Modern Physics Letters A,
34(29):1950240, 2019.

\bibitem{vv}Pawel O Mazur and Emil Mottola. Gravitational vacuum condensate stars.
Proceedings of the National Academy of Sciences, 101(26):9545–9550, 2004.

\bibitem{ww} David F Mota. On the spherical collapse model in dark energy cosmologies, Astronomy \& Astrophysics, 421(1):71–81, 2004.
\bibitem{xx} Rong-Gen Cai and Anzhong Wang. Black hole formation from collapsing
dust fluid in a background of dark energy. Physical Review D, 73(6):063005, 2006.
\bibitem{yy} Pedro G Ferreira and Michael Joyce. Structure formation with a self-tuning scalar field. Physical Review Letters, 79(24):4740, 1997.

\bibitem{zz} Soma Nath, Subenoy Chakraborty, and Ujjal Debnath. Gravitational collapse due to dark matter and dark energy in the braneworld scenario. International Journal of Modern Physics D, 15(08):1225–1236, 2006

\bibitem{ab}. Hasrat Hussain Shah, Farook Rahaman, Amna Ali, and Sabirudin Molla.
Gravitational collapse of an interacting vacuum energy density with an anisotropic fluid. Physics of the Dark Universe, 24:100291, 2019.

\bibitem{ac} Eugeny Babichev, Vyacheslav Dokuchaev, and Yu Eroshenko. Black hole
mass decreasing due to phantom energy accretion. Physical Review Letters, 93(2):021102, 2004.

\bibitem{ad}Caldwell, R.R. , Kamionkowski, M. , and Weinberg, N.N. (2003). Phantum Energy: Dark Energy with $w<-1$ Causes a cosmic Doomsday. Physical Review Letters, 91(7). doi.1103/physervlett91.071301  

\bibitem{ae}Caldwell, R. . (2002). A phantom menace? Cosmological consequences of a dark energy component with super-negative equation of state. Physics Letters B, 545(1-2), 23–29. doi:10.1016/s0370-2693(02)02589-3 



\bibitem{af} V.Sahni, "Dark Matter and Dark Energy," arXiv:astro-ph/0403324(2004).
\bibitem{ag} V. Sahni, "Cosmological Surprises from Braneworld models of Dark Energy," arXiv:astro-ph/0502032(2005).
\bibitem{ah} Ma, C.-P., Caldwell, R. R., Bode, P., \& Wang, L. (1999). The Mass Power Spectrum in Quintessence Cosmological Models. The Astrophysical Journal, 521(1), L1–L4. doi:10.1086/312183 .


\bibitem{ai}Chakraborty, S., \& Bandyopadhayaya, T. (2008). Collapsing Inhomogeneous Dust In The Background Of Perfect (Or Anisotropic) Fluid. International Journal of Modern Physics D, 17(08), 1271–1281. doi:10.1142/s0218271808012814 
\bibitem{aj} LI, Z.-H., \& WANG, A. (2007). Existence Of Black Holes In Friedmann–Robertson–Walker Universe Dominated By Dark Energy. Modern Physics Letters A, 22(22), 1663–1676. doi:10.1142/s0217732307024048  
\bibitem{ak} H.H. Shah, F.Rahaman. A.Ali and S. Molla, Gravitational Collapse of an interacting vacumm energy density with an anisotropic fluid, Physics of the dark Universe, 24:100241(2019). 
\bibitem{al}Chan, R., da Silva, M. F. A.,\& Villas da Rocha, J. F. (2009). Star models with dark energy. General Relativity and Gravitation, 41(8), 1835–1851. doi:10.1007/s10714-008-0755-9 .
\bibitem{am}Cai, R.-G., \& Wang, A. (2005). Cosmology with interaction between phantom dark energy and dark matter and the coincidence problem. Journal of Cosmology and Astroparticle Physics, 2005(03), 002–002. doi:10.1088/1475-7516/2005/03/002 
\bibitem{an}Shah, H. H., \& Iqbal, Q. (2017). Gravitational collapse of dark matter interacting with dark energy: Black hole formation. International Journal of Modern Physics D, 26(13), 1750142. doi:10.1142/s0218271817501425 .




\end{thebibliography}
\end{document}